\documentstyle[11pt,paspconf,epsf]{article}

\begin{document}

\title{ SEARCH FOR EXTRA-SOLAR PLANETS THROUGH MONITORING
MICROLENSING EVENTS FROM ANTARCTICA}
\author{Kailash C. Sahu}
\affil{Space Telescope Science Institute, 3700 San Martin Drive, 
Baltimore, MD 21218}

\begin{abstract}

During the months when the galactic bulge is 
visible from the southern hemisphere, there are typically about 8 to 10
on-going microlensing events at any given time.
 If the lensing stars have planets around 
them, then the signature of the planets can be seen as sharp, 
extra peaks on the microlensing light curves. And if every lensing star has a Jupiter around it, then the probability of detecting an extra spike
is of the order of 10\%. 
Thus continuous and frequent monitoring of the on-going microlensing events,
with a sampling interval of a few hours, provides a powerful new 
method to search for planets around lensing stars.
 
Such monitoring programs are now being carried out using a network of
1-meter class telescopes situated at appropriately spaced longitudes in the southern hemisphere (for example, by PLANET collaboration).
However, the galactic bulge is visible from the south pole throughout this
period, and hence a single automated telescope at the south-pole can provide
an efficient means of carrying out the monitoring program.
Up to about 20 events can be monitored during a single 3-month season
with a 1-meter telescope, potentially leading to the detection of two 
planetary signals. The telescope can also be used for several
other research projects involving microlensing and variability of stars.

\end{abstract}


\section{Power of gravitational lensing}

The last time a planet was discovered by its direct light
was about 2 centuries ago when Sir William Herschel
saw the planet Uranus through his telescope in 1791.
As described below, (almost) all subsequent planet discoveries 
have been made {\it not} by the
direct light of the planet but by its gravitational effect. 
 
In 1845, the French astronomer Leverrier predicted the position
of Neptune from the orbital perturbations of Uranus. 
The prediction was then observationally followed up by Johan Galle, who
discovered Neptune in a single night of observations.
In 1930, when Lowell predicted
the position of Pluto from the orbital perturbations of Neptune, which
was then easily discovered by Tombaugh. The discovery of the first 
definitive extra-solar planet around the pulsar PSR1257+12 was again 
through the gravitational effect of the planet (Wolszczan and Frail, 1992).
The most recent flurry of discoveries of planets around 
the nearby stars have made use of the gravitational effect 
in a different facet, namely the radial velocity 
perturbation it causes on the parent star (Mayor et al. 1995;
Marcy and Butler, 1995).
On the other hand, a tremendous amount of
effort has been spent in looking for planets around 
other stars though other esoteric means, such as spatial
interferometry or adaptive optics. 
While some of these efforts will no doubt bear fruit in the 
near future as we overcome the technical challenges they pose, 
they have borne very little fruit so far.
The reason is not difficult to understand: the gravitational effect,
in almost all cases, makes use of the bright nearby object whereas
the other methods seek to overcome the effect of the bright nearby
object through technology. In the case of spatial
interferometry or adaptive optics, one must always fight
to keep the light of the bright star down in order to
detect the faint planetary signal in the presence of this
highly dominant bright source. In other words, the bright star
always acts as  a hindrance to the search, 
and is always something that one must win over in order to
be able to detect the much fainter planet nearby.
The situation is reversed in case of the gravitational effect of the
planet, in which case, one simply uses the features in the brighter
object to look for perturbations.
In case of Neptune and Pluto, the nearby brighter object was used 
to look for perturbations in its orbit. In case of the pulsar PSR1257+12,
the pulse period distribution of the pulsar itself was used to 
look for the effect due to the planet.
And in case of the radial velocity measurements, 
the absorption lines from the parent star was necessary to
look for the effect of the planet.  

The technique  microlensing also
uses the brighter object nearby, and the star in this case too helps in the search for the planet nearby.  This may potentially be a very
powerful tool to look for extra-solar planets, and 
as discussed in
more detail later, this is the only method sensitive to the search 
for Earth-like planets around normal stars, using ground based 
observations. Furthermore, this is the only method which can provide a 
statistics on the masses and orbital radii of extra-solar planets.
It must be noted however that microlensing does have its
selection effects, and this method is more
sensitive to detection of planets around {\it low} mass stars
since, statistically, a large fraction of the lenses are expected to be
low mass stars.

The search for extra-solar planets through microlensing 
requires frequent and continuous monitoring of on-going
microlensing events, and this paper describes how a telescope
in Antarctica can be an ideal instrument for such a pursuit.
 
\section{Microlensing due to stars}   

The idea of microlensing by stars is not new. In 1936, Einstein wrote a
small paper in {\it Science}  
where, he did `a little calculation' at the request of his friend
Mandal and showed that if a star happens to pass very close to another 
star in the line of sight, then the background star will be lensed
(Einstein, 1936). However, he also 
dismissed the idea as only a theoretical exercise and remarked that 
there was `no hope of observing such a phenomenon directly'. He was 
right at that time; the probability of
observing is  less than one in a million, and with the 
technology of 1936, there was no way one could observe this directly.

Paczy\'nski, in two papers written in 1986 and 1991, noted that if one 
could monitor a 
few million stars, one could observe microlensing events, perhaps
as a signature of the dark matter towards the LMC, or by known stars 
towards the Galactic Bulge (Paczy\'nski, 1986; Paczy\'nski, 1991;
also see Griest, 1991). 
The project was taken up immediately by 
three groups and the first observed microlensing event was reported 
towards the LMC in 1993. By now, more than 100 events have been
discovered, mostly towards the Galactic Bulge.

Out of the more than 100 microlensing events detected so far, only
8 are observed towards the LMC, the rest overwhelming
majority being towards the Galactic Bulge. 
There is, as yet, no general consensus on the nature and location of the
lenses towards the LMC, and the lenses could be anywhere between the local disk
(Gould et al. 1994) or the halo (Alcock et al. 1995) and the LMC itself
(Sahu 1994a,b; Wu, 1994). Towards the galactic bulge however, 
the general consensus is that a majority of the lenses are
stars in the line of sight.

It is then a logical step to look for planets around these 
lensing stars through microlensing: the rest of this paper
deals with the details of such a method to search for extra-solar
planets.

\section{Theoretical Aspects of Microlensing}

Before proceeding into the details of the lensing due to binaries and planets,
it is useful to review the basics of the lensing by a single star.  
For the details of the theoretical aspects of the lensing
by a star, the reader may refer to the excellent review
article by Paczy\'nski (1996) and the very exhaustive 
monograph devoted to the subject of Gravitational Lensing
by Schneider, Ehlers and Falco (1992). The basic information
which we will need later are essentially the following.

\begin{figure}
\centering \leavevmode
\epsfxsize=9truecm \epsfbox{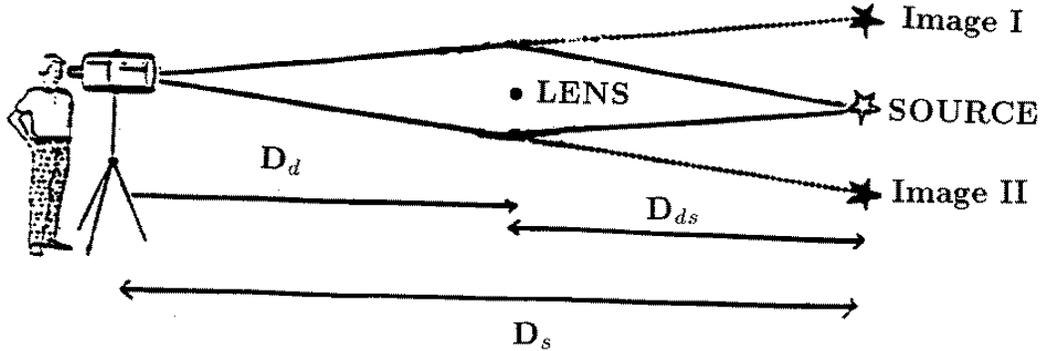}
\caption{Schematic geometry of the gravitational microlensing. 
The presence of the lens causes the image of the source to split into 
two, their combined brightness being always larger than that
of the unlensed image. Note that the deflection
due to the lens and the separation of the images are greatly
exaggerated in this schematic diagram.}
\end{figure}
 
With the lensing geometry as described in Figure 1, the
Einstein ring radius R$_E$ can be written as
$$ R_E^2 = {{4GMD}\over {c^2}}, D = {D_{ds} D_d\over D_s}  \eqno (1)$$
 
\noindent where M is the  the mass of the lensing object,
D$_d$ is the the distance to the lensing object, 
D$_{ds}$ is the 
distance from the lens to the source, and
D$_s$ is the distance from the observer to the source.

The amplification due to the microlensing depends only on the
impact parameter, which can be written as 

$$A =  {{u^2+2}\over u(u^2+4)^{1/2}} \eqno (2)$$
where $u$ is the impact parameter in units of $R_E$.
 
This equation can be easily inverted to
derive the impact parameter from a given amplification
$$ u = 2^{1/2} [A (A^2 - 1)^{-1/2} -1]^{1/2} \eqno (3)$$
which can be used to derive the minimum impact parameter
$u_m$ from an observed light curve.

 \begin{figure}
\centering \leavevmode
\epsfxsize=10truecm \epsfbox{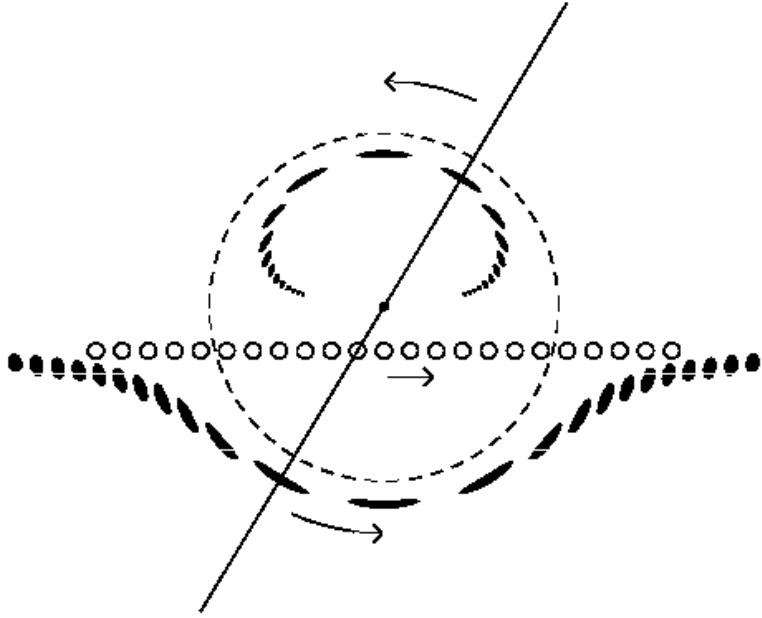}
\caption{ This figure shows how the
apparent positions and the sizes of the images change at various stages of the
microlensing. In this geometry the position of the lens, 
indicated by a solid dot,  is fixed, 
and the open circles show the actual positions of the source.
The filled circles show the images of the source as the source passes
close to the lens in the plane of the sky. The dashed circle is
the Einstein ring of the lens. 
At any instant, the source, the lens and the two images lie on a straight 
line.  (Taken from Paczy\'nski, 1996)}
\end{figure}

The the time scale of microlensing is the time taken
by the source to cross the Einstein ring radius, which 
is given by
  
$$ t_0 = {R_E \over{V_e}}  \eqno (4)$$
where $V_e$ is the tangential velocity of the lensing object.
The impact parameter
at any time during the microlensing event can be expressed as
$$ u = [u_m^2 + ({t-t_{m}\over t_0})^{2}]^{1/2} \eqno (5)$$
where $t_{m}$ is the time corresponding to  the minimum impact parameter 
 (or the maximum amplification).
 
From Eq. 1 and 4, the mass of the lens can be expressed as
$$M = {[t V_e c]^2 \over {4GD}} \eqno (6)$$

\begin{figure}
\centering \leavevmode
\epsfxsize=10truecm \epsfbox{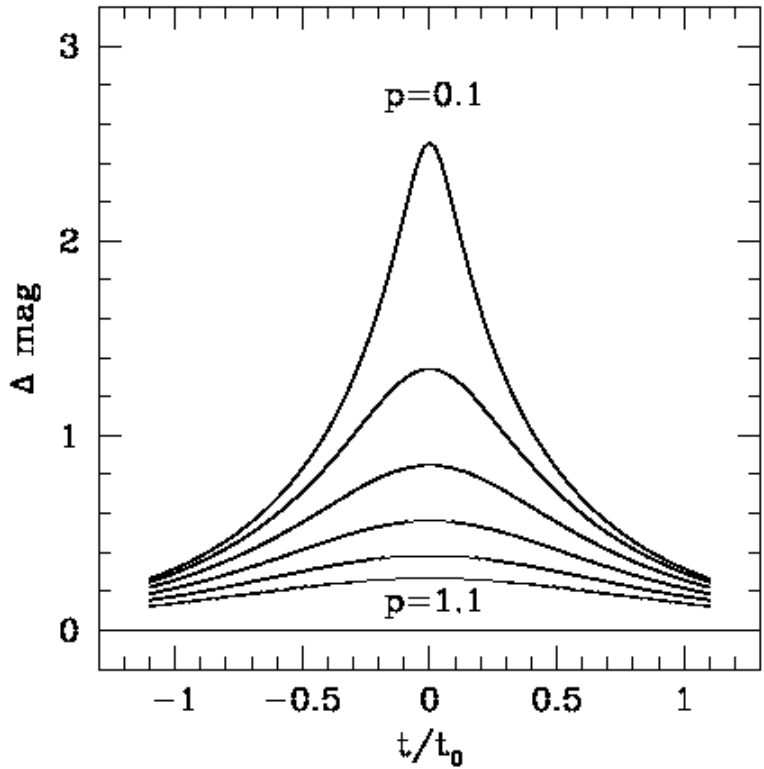}
\caption{The microlensing light curves as a function of impact parameter. }
\end{figure}

\subsection{Effect of Extended Source}

In the case of the microlensing events towards the LMC and the Galactic Bulge,
the point-source approximation may not always be valid. This is particularly 
the case if the LMC events are caused by the LMC stars and the Bulge events 
are caused by the Bulge stars, in which case the distance between the source 
and the the lens is not large. Consequently, the Enstin ring radius is smaller,
in which case the source size cannot be neglected. (For more details 
see Sahu, 1994b, Sahu 1997). This is also very important for lensing caused by 
planetary mass objects since the Einstein ring radius of a planet may not 
always be much larger than the size of the source. 
In such a case, different parts of the source will be amplified differently and the net amplification can be expressed as (Eq. 6.81 of Schneider,
Ehlers and Falco, 1992)
$${\int d^2y \ I(y) \ \mu_p(y) \over \int d^2y \ I(y)} \eqno (7)$$
\noindent where I(y) is the surface brightness profile of the source,
$\mu_p(y)$ is the amplification of a point source at point
$y$, and the integration is carried out over the entire surface
of the source.

In extended-source approximation, since different parts of the source 
are amplified differently, the limb darkening effect can be important. 
This can make the event chromatic and the ratios of the emission/absorption
in the star features in the source star can vary during the event (Loeb and Sasselov, 1995). Such effects have 
indeed been seen in case of MACHO 95-30 (Alcock et al. 1997).
The extended source effect can be particularly important
in case of planetary events where, in general, the source-size
cannot be neglected.  

If the source can be approximated as a disk of uniform brightness, then the
maximum amplification, when the source and the lens are perfectly aligned, is given by

$$ A_{max} = [1 + {4R_E^2\over{r_{0}^2}}]^{1\over{2}} \eqno (8)$$
\noindent where $r_0$ is the radius of the source. When the Einstein
ring radius is the same as the radius of the source, the maximum 
possible amplification in such a case is $\sim$2.24.

\section{Planets as lenses}

\subsection{Observational characteristics} 

The light curve due to a binary lens, unlike the single lens, can be complex 
and can be very different from the mere superposition
of two point lens light curves.
In case of a double lens, the lens equation, which is 
a second order equation for a single lens, becomes
two 5th order equations (or one 5th order equation in the complex plane,
Witt and Mao, 1995).  
The most important new feature is the formation of 
caustics, where the amplification is infinite for a point source,
but finite for a finite size source. When the source crosses
a caustic, an extra pair of images forms or disappears. 
A full description of the microlensing due to a 
double lens is given by Schneider and Weiss (1986).

If the lensing star has a planetary 
system, the effect of the planet on the microlensing 
light curve can be treated as that of
a binary lens system. The signature of the planet can be seen,
in most cases, as sharp extra
peaks in the microlensing light curve. Computer codes for analysis of
such data have been developed by Mao and Di Stifano (1995)
and Dominik (1996).

\begin{figure}
\centering \leavevmode
\epsfxsize=11truecm \epsfbox{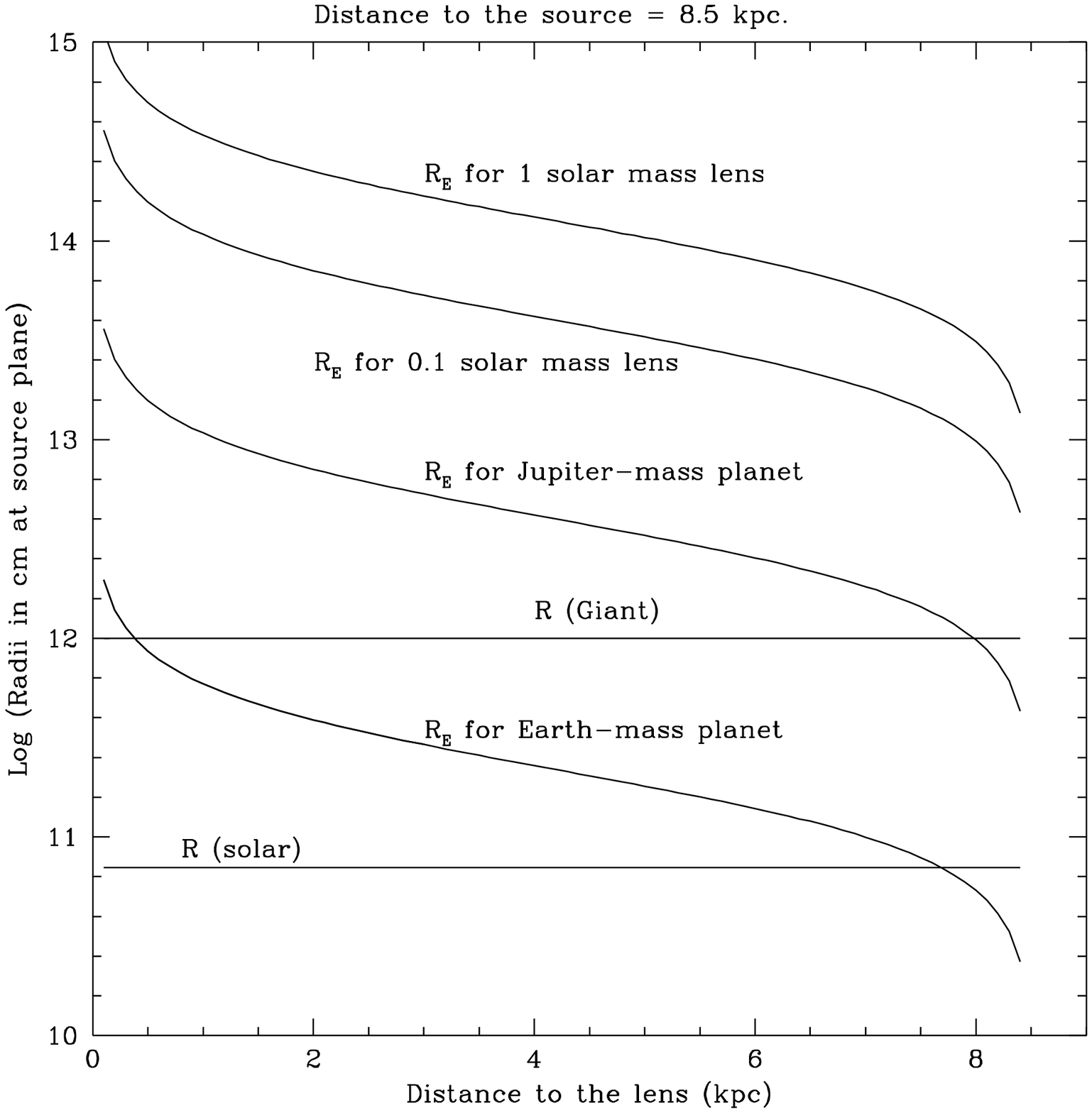}
\caption{The figure shows the sizes of the Einstein ring radii R$_E$ at the 
source plane,
for a lensing event towards the Galactic Bulge. R$_E$
for an Earth-mass planet a solar mass star are shown.
Also shown are the actual radii of a solar type star and a typical giant star
as projected onto the source plane, which are denoted by R(solar) and R(giant)
respectively. For a Jupiter-mass planet, R$_E$ is
almost always larger than the radius of a giant star, so the effect 
of Jupiter can always be significant.  But for an Earth-mass planet, 
there is only a small 
parameter space where the Einstein ring radius is larger than the size
of a giant star. 
If the source is  a main-sequence star like our Sun, the
Einstein ring radius due to an earth-mass planet is almost always larger 
than the source size, and hence the amplification can be large.}
\end{figure}

\begin{figure}
\centering \leavevmode
\epsfxsize=11truecm \epsfbox{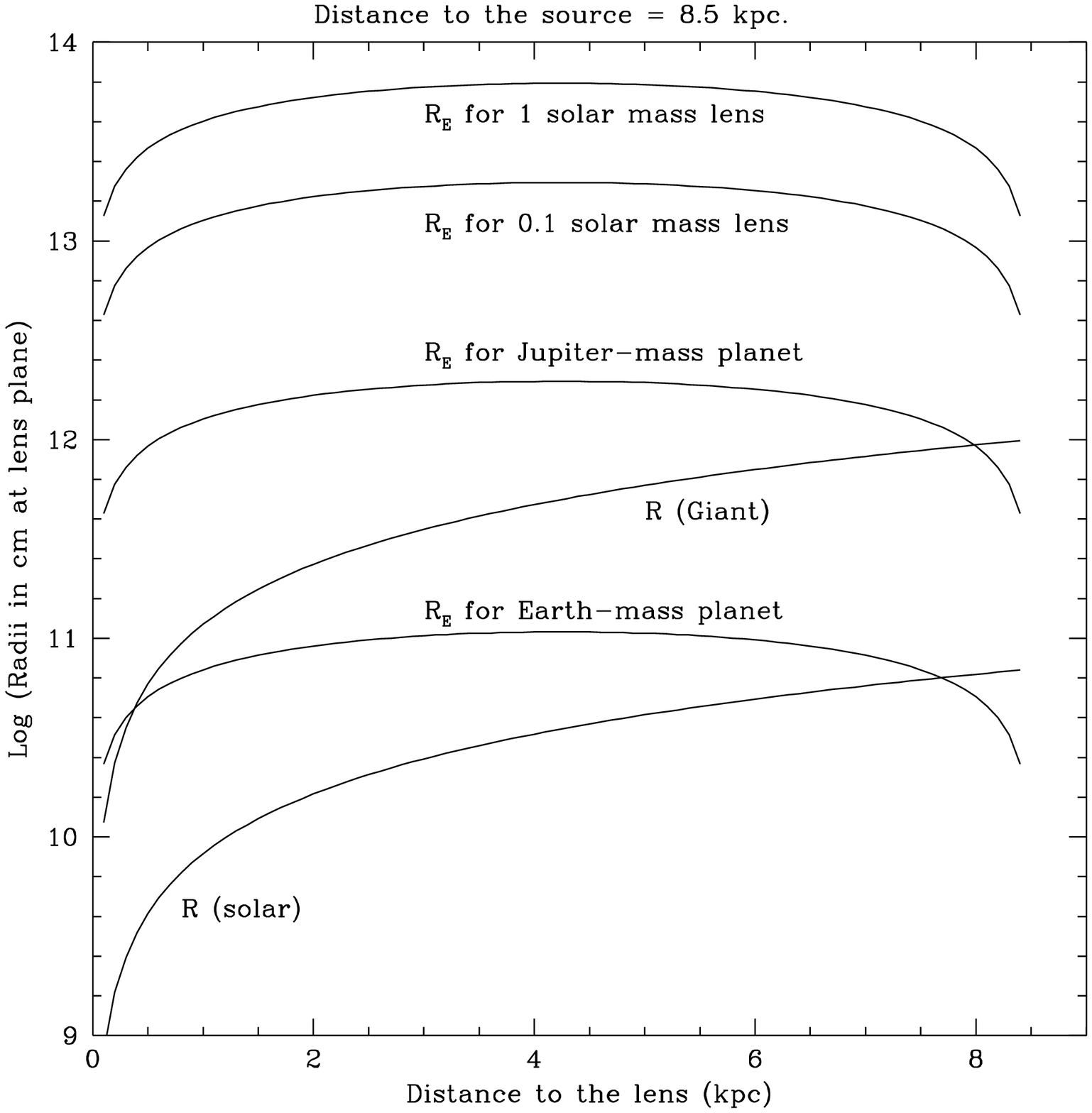}
\caption{ The same as Fig. 4, but here the sizes of the Einstein ring 
radii and the sizes of the sources are all as seen projected onto the 
lens plane.}
\end{figure}

 It was first shown by Mao and Paczy\'nski (1991) that about ~10\% of the 
lensing events should show the binary nature of the lens,
and this effect is strong even if the companion is a planet.
The problem of microlensing by a star with a planetary system 
towards the Galactic Bulge was further
investigated by Gould and Loeb (1992). They noted that, for a 
solar-like system  half way 
between us and the Galactic Bulge, Jupiter's orbital radius coincides
with the Einstein ring radius of a solar-mass star.
Such a case is termed `resonant lensing' which increases the
probability of detecting the planetary signal. In $\sim$20\% of the cases, 
there would be a signature with magnification larger than 5\%.
 
The importance of  the resonant lensing can be qualitatively 
understood as follows. 
In Fig. 2, the impact parameter changes through a large range as the
source passes close to the lens. The positions of two images
formed by the lensing effect change continuously, but they remain close 
to the Einstein ring for a large range of impact parameters. 
So, the effect of the planet can be large if the planet happens to be 
close to the Einstein ring, which causes a further amplification. This also
qualitatively explains why the probability of observing the
effect of the planet increases if it is close to the 
Einstien ring. 
 
In a large number of cases however, the resulting light curve
due to a planet plus star system is close to the superposition of
two point lens light curves (Fig. 6). This is particularly true
when the star-planet distance is much larger than R$_E$.
In such a case, the time scale 
of the extra peak due to the planet, $t_p$, and the time scale of the
primary peak due to the star, $t_s$, are related through the relation
$t_p / t_s = \sqrt{(m_p/M_s)}$ where $m_p$ is the mass of the planet and $M_s$
is the mass of the star. Furthermore, if the source size cannot be
neglected, the maximum amplification
given by Eq. 7 remains valid. Fig. 4 and 5 show the sizes of the
Einstein ring radii due to planetary and stellar mass lenses as a function of
distance to the lens. 
The typical sizes of the main-sequence and giant sources
are also shown. In such a case, it is clear that  
the size of the source is almost always smaller than the Einstein
ring radius of a Jupiter-mass planet, as a result
the amplification due to the planet can be large.
The amplification can also be
large for an Earth-mass planet if the source is a main sequence star.
However, if the source is a giant-type star, then there
is only a fixed range of D$_d$ where the amplification due to
an Earth size planet can be large.

In general. the situation is different in case of formation of caustics.
Fig. 5 shows the effect of the formation of caustics
and consequent high amplification and sharp peaks
caused by the planets. The planets, in this case, are
situated at different orbital radii and the mass of each planet 
is 10$^{-3}$ times that of the primary. 
The solid curve shows the light curve without the presence of the
planets. The dashed and the dotted curves correspond to two 
representative tracks of the source {\it with} the presence of the
planets (Wambsganss, 1997). 

\begin{figure}
\centering \leavevmode
\epsfxsize=9truecm \epsfbox{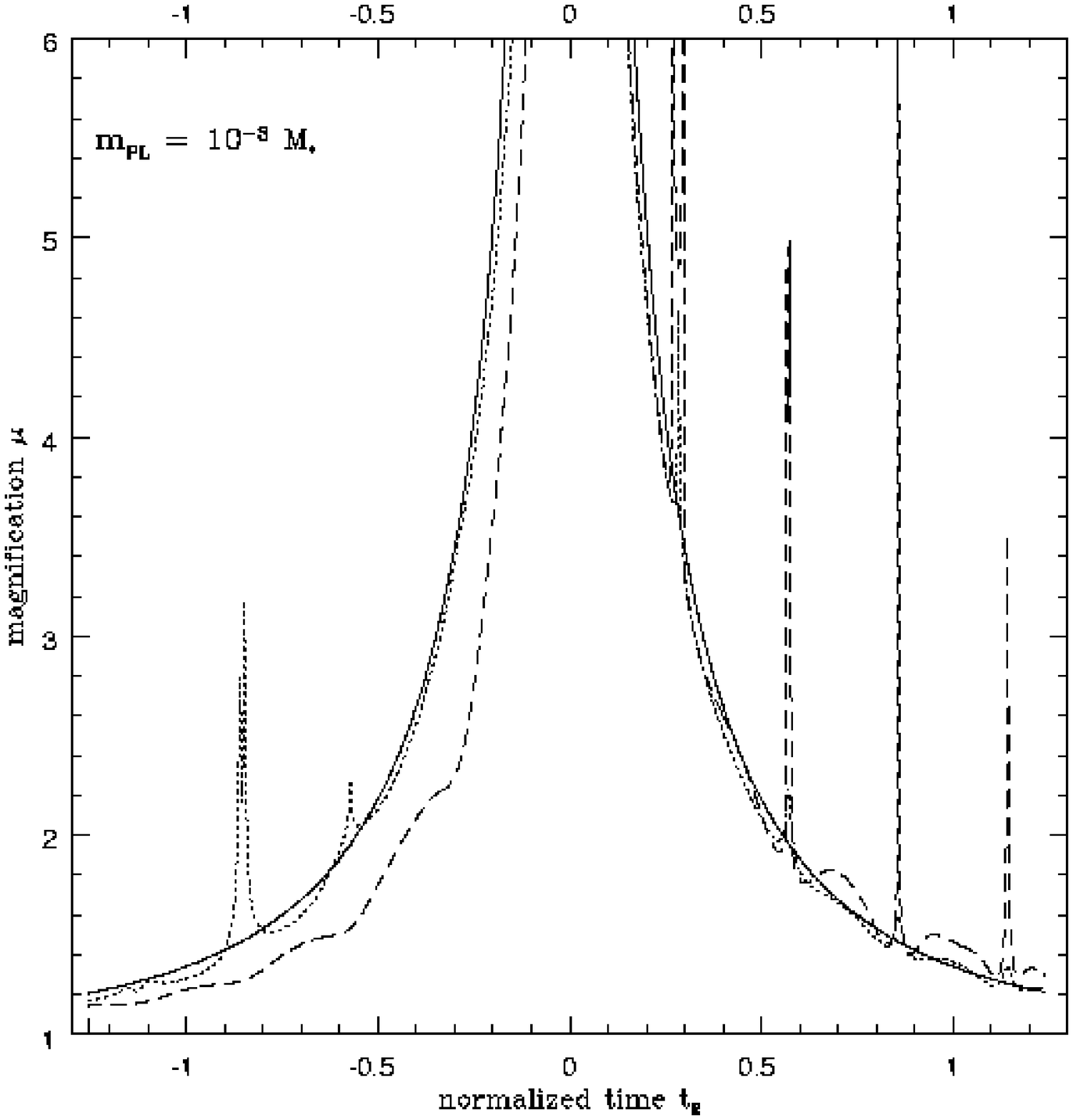}
\caption{Effect of a few planets situated at different orbital radii on the
microlensing light curve. The mass of each planet is 10$^{-3}$ times that 
of the primary. The projected distances of the planets from the primary,
in units of R$_E$, are 0.57, 0.65, 0.74, 0.86, 1.16, 1.34, 1.55 and 1.77.
The solid curve shows the light curve without the presence of the
planets. The dashed and the dotted curves correspond to two 
representative tracks of the source {\it with} the presence of the
planets. This shows how the influence of the planet on the
microlensing light curve can be both upward and downward,
and illustrates the possible high amplification and sharpness of the
extra peaks caused by the planets (taken from Wambsganss, 1997).}
\end{figure}

The minimum duration of the extra feature due to the planet, to a first 
approximation, is the time taken by the source to 
cross the caustic, which can be about 1.5 to 5 hrs. The maximum duration 
of the spike is roughly the time taken by the planet to cross its own Einstein ring.
Using a reasonable set of parameters (the lower mass of the planet is
taken as that of the Earth, the higher mass is assumed to be that of Jupiter)
this can be a few hours to about 3 days.  Any follow-up program must be accordingly adjusted so that the extra
feature due to the planet is well sampled.
 
{\subsection{Theoretical work}}
 
A full description of the theoretical aspects of planets acting as lenses
is beyond the scope of this review. 
To date, there are a few countable number of papers which deals with the 
theoretical prediction of planetary signals on the light curve,
which the reader may refer to (Bolatto and Falco, 1995; Bennett and 
Rhie, 1996; Wambsganss, 1996; and Peale, 1997). 

\section{Requirements for a Follow-up Monitoring Program}

The first requirement for a follow-up monitoring program
is access to the `alert' 
events. With the alert capability of the survey programs firmly in place, 
it is now possible
to build a follow-up program. At present, the alert events from the MACHO collaboration at a given time is sufficient to carry out a ground based 
follow up program with small telescopes towards the Galactic Bulge. 
After OGLE and EROS II experiments have their alert systems in place, 
the number of on-going alert events at a given time will increase and
it may be possible to extend such follow-up networks to larger telescopes, 
and also perhaps towards the LMC.

The second requirement is the ability to monitor hourly. It should be 
noted that, assuming that the longer time scale events are mostly due to slower 
proper motion of the lensing star, the time scale of the planetary signal 
approximately scales with the time scale of the main event. So the monitoring, 
in general, can  be less frequent for longer time scale
events.  
 But typically, as noted before, the time scales of the planetary event 
can be a few hours to a few days. So the follow-up monitoring 
program must have the capability to do hourly monitoring 
so that the extra feature due to the planet
is well sampled. For discrimination against any other short term variations,
some color information is also useful, since the microlensing is expected 
to be achromatic, where as most other types of variations 
are expected to have some chromaticity. Thus it is preferable to have 
a few observations in two colors. 

The third requirement is to have 24-hour coverage in 
the monitoring program. This calls for telescopes at appropriately spaced 
longitudes around the globe.
 
\section{A case for an automated  telescope in Antarctica}

An automated telescope close to the south pole fits the bill perfectly since it
not only satisfies all the above requirements, but also has the advantage that
it can be operated remotely. 

\begin{figure}
\centering \leavevmode
\epsfxsize=12truecm \epsfbox{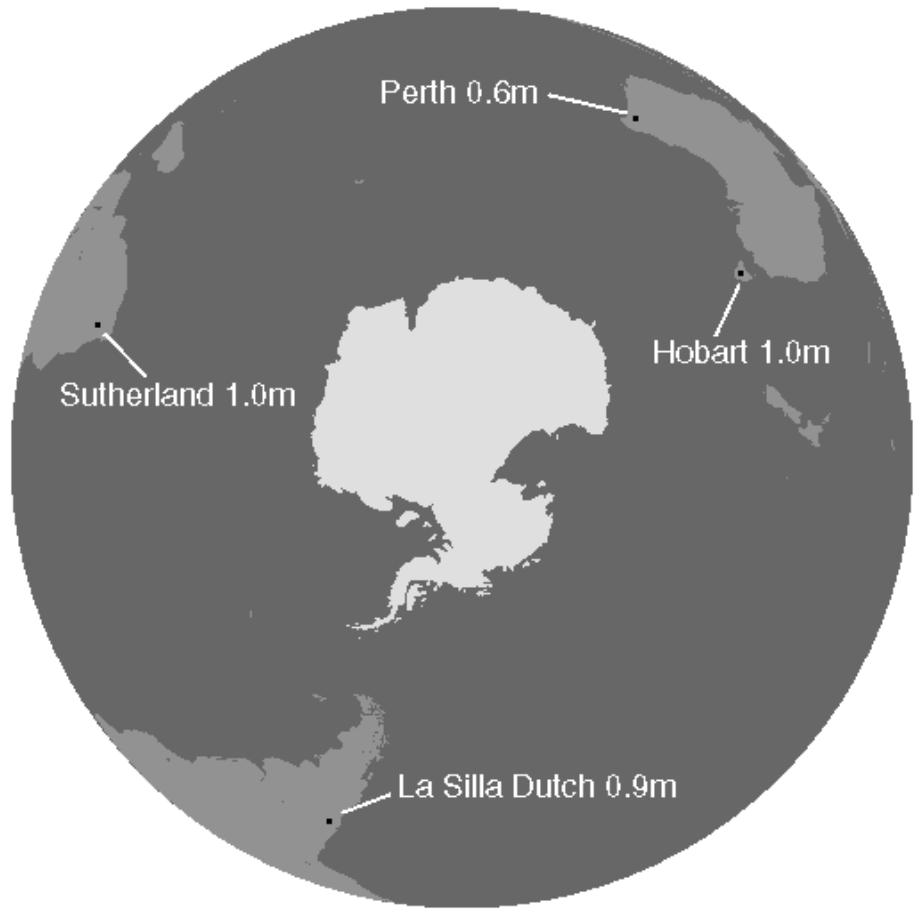}
\caption{The location of the telescopes used by the PLANET
collaboration for follow-up monitoring program of 
microlensing events (courtesy of PLANET collaboration). 
A telescope in Antarctica can achieve the
same goal of 3 telescopes for 24-hour coverage in the monitoring program. }
\end{figure}

The PLANET collaboration currently uses 3 or 4 telescopes situated
at appropriately spaced longitudes in the southern hemi-sphere
to achieve a near-continuous coverage in the monitoring program
(Albrow et al. 1995, 1996a,b; 1997).
The same can be achieved with a single telescope in Antractica (Fig. 6).

At any given time during the $\sim$3-month period when the galactic bulge
is observable, there are typically 8 to 10 on-going microlensing
events. This number is likely to increase after
the EROS II and OGLE II survey programs  have their alert systems fully operational.

With a 1-meter size telescope, about 7 events can be monitored continuously
with  a sampling interval of 2 to 5 hours required for this project.
Taking the typical monitoring duration (which is larger than the microlensing
time-scale) to be  1-month, about 20 events can be monitored in a single
3-month period during which the galactic bulge is visible. Since the 
probability of detection is about 10\% if every star has a Jupiter,
 that should lead to the detection of 2 planetary
events in a single season. Even non-detection can provide very strong
constraints on the presence of planets around lensing stars.
With a bigger telescope, a  larger number of events can be monitored, 
with a correspondingly larger number of expected planetary detections.

 Once the number of events towards the LMC and SMC increase, 
as indeed expected in the near future after the EROS II and OGLE 
survey programs are fully operational,  frequent monitoring of
ongoing LMC and SMC microlensing events is a promising possibility.
This would make the search strategy sensitive to a very different region in the
orbital parameter space, and to a very different group of stars.
 Both the LMC and the SMC being circum-polar 
objects, such an extension of the monitoring program
will keep the telescope busy for most of the year.

\section{Spin-off and other scientific projects}
 
Although the primary goal would be  to search for 
extra-solar planetary systems,  numerous other interesting 
projects can be done with such a telescope, some of which are
 briefly described below. Most of these 
projects (1 to 3 below) can be done with the same data and will not 
need any additional observations. 

{\subsection{Short time-scale and new microlensing events}

 Since the line of sight towards the galactic bulge passes through the dense
concentration of stars in the galactic disk, the optical depth to 
microlensing in the direction of the bulge is known to be high -- ~ 1 x 10$^{-6}$. If small mass stars (M $<$ 0.1 M$_\odot$) contribute 
significantly to the
mass budget of the galaxy, then a sizable fraction of this optical depth
would be due to these small mass stars. 

The Einsten ring radius of a 0.01 M$_\odot$ object for a source-lens
distance of 1 kpc is about 3 x 10$^{12}$ cm  (its variation 
with distance is given by Eq. 1.) Assuming 
a typical proper motion of about 200 km/s, the time scale for such events 
would be about 2 days (about 16 hours for m =  0.001 M$_\odot$ stars).
The current microlensing survey programs are insensitive to such
short time-scale events
and such events would essentially go unnoticed by the survey 
programs. However, such events can be easily detected by the monitoring
program mentioned here with the automated telescope.

There would be  typically about 20,000 stars per one 
1000 X 1000 pixel CCD image. Thus  about 2 $\times$ 10$^6$ stars would be observed in about 5 frames. Thus, if 50\%  of the mass is in form of small 
mass stars of mass m=0.01 M$_\odot$,  1 such event would be detected every 2 
nights (or about 1 event every 16 hours if m=0.001 M$_\odot$).

Thus, this program can provide an answer to the contribution of small 
mass stars to the mass budget of the galaxy. Even non-detection will provide 
a strong upper limit to their contribution.

In addition, the telescope can be used for detecting new microlensing events.
This can perhaps be best done in the near-IR wavelengths where the low
extinction allows one to probe deeper into the galactic bulge.
The high IR transmissivity near the south pole makes such a project
particularly attractive.

{\subsection{Binary Events}}

 The resolution of sharp caustic crossing in binary events, which 
may last only a few hours, can be clearly detected  from such a monitoring
program (Mao \& Di Stefano 1995).  
 This can be used to get the binary fraction and characteristics
of the lensing objects.
 
{\subsection{Extended sources}}

High-amplification lensing of giants can resolve the
structure of the source star. Such an extended source effect 
can produce chromatic effects that can be 
quantified through high-quality multi-band photometry near the event peak.
Thus the observations can be used to `map' the source star.

{\subsection{Variable stars}}

 Finally, the resulting  data base of frequently-sampled bulge fields, each 
containing $>$10,000 stars would be clearly useful for the study of 
variable stars.  In addition, the telescope can  be used for the
purpose of studying some specific variable  stars.
In particular, the programs which need continuous monitoring for more than 10 hours (e.g. the study of stellar oscillations)
can greatly benefit from such a telescope.

\acknowledgments

I would like to thank all the members of the PLANET collaboration, for their
help and co-operation in all our attempts on the folow-up monitoring of microlensing events.


\begin{references}
\reference Albrow, M., et al. (PLANET collaboration), 1995, in Proc.
IAU Symp. 173, Eds. C.S. Kochanek, J.N. Hewitt, p227
\reference Albrow, M., et al., 1996a, Proc. 12th IAP Conference, in press (astro-ph/9610128) 
\reference Albrow, M., et al., 1996b, BAAS, {\bf 27}, 1449 
\reference Albrow, M. et al., 1997, Proc. STScI Symp. on {\it Planets beyond the solar system and the next 
generation of space missions}, Ed. D. Soderblom, ASP Conf. Series.
 \reference Alcock, C. et al., 1995, ApJ, {\bf 454}, L125
\reference Alcock et al., 1997, Preprint (astro-ph/9702199)
\reference Bahcall, J.N., Flynn, C., Gould, A., Kirhakos, S., 1994,
ApJ, {\bf 435}, L51
  \reference Bennett D.P, Rhie, S., 1996, ApJ, {\bf 472}, 660
 \reference Bolatto, A.D., Falco, E.E., 1994, ApJ, {\bf 436}, 112 
\reference Dominik, M., 1996, Ph.D. Thesis, Univ. Dortmund
\reference Einstein, A., 1936, Science, {\bf 84}, 506
 \reference Gould, A., Loeb, A.,  1992, ApJ, {\bf 396}, 104
\reference Griest, K., 1991, ApJ, {\bf 366}, 412
\reference Gray, D., 1997, Nature, {\bf 385}, 795
\reference Kiraga, M, Paczy\'nski, B., 1994, ApJ, {\bf 430}, 101
 \reference Loeb, A., Sasselov, D., 1995, ApJ, {\bf 491}, L33
 \reference Mao, S., Paczy\'nski, B., 1991, ApJ, {\bf 374}, L37
\reference Mao, S., Di Stifano, R., 1995, ApJ, {\bf 440}, 22 
\reference Marcy, G.W., Butler, R.P., 1996, ApJ, {\bf 464}, L153
\reference Mayor, M., Queloz, D.A., 1995, Nature, {\bf 378}, 355
 \reference   Paczy\'nski, B., 1986, ApJ, {\bf 304}, 1
\reference   Paczy\'nski, B., 1991, ApJ, {\bf 371}, L63
\reference   Paczy\'nski, B., 1996, Ann. Rev. Astron. Astrophys. {\bf 34}, 415
\reference Peale, S.J., 1997, Icarus, {\bf 127}, 269
\reference   Sahu, K.C., 1994a, Nature, {\bf 370}, 275
\reference  Sahu, K.C., 1994b, Pub. Astron. Soc. Pac., {\bf 106}, 942
\reference Sahu, K.C., 1997, ``Detecting planets through microlensing",
Proc. STScI Symp. on {\it Planets beyond the solar system and the next 
generation of space missions}, Ed. D. Soderblom, ASP Conf. Series.
\reference Schneider, P., Ehlers, J. \& Falco, E.E., 1992, ``Gravitational 
Lensing", (Springer-Verlag).
\reference Schneider, P., Weiss, A., 1986, Astron. Astrophys. {\bf 164}, 237
 \reference Wambsganss, J., 1997, MNRAS, {\bf 284}, 475 
\reference Witt, H., Mao, S., 1994, ApJ., {\bf 430}, 505
\reference Wolszczan, A., Frail, D.A., 1992, Nature, {\bf 355}, 145
\reference Wu, X-P., 1994, ApJ., {\bf 435}, 66
 \end{references}
\end{document}